\documentclass[conference]{IEEEtran}
\usepackage{multirow}
\usepackage{multicol}
\usepackage{lipsum}
\usepackage{graphicx}
\usepackage{graphics}
\usepackage{amsmath}
\usepackage{amsbsy}
\usepackage{blindtext}
\usepackage{tcolorbox}
\usepackage{amssymb}
\usepackage{scalerel}
\usepackage{mathabx}
\usepackage{verbatim}
\usepackage{booktabs}
\usepackage{rotating,tabularx}
\usepackage{mathrsfs} 
\usepackage{hyperref}

\usepackage{adjustbox}
\usepackage{soul}
\usepackage{xcolor}
\usepackage{cite}
\usepackage{tikz}
\usepackage{color, colortbl}
\usepackage[first=0,last=9]{lcg}
\definecolor{LightGray}{rgb}{0.7,0.7,0.7}

\usepackage[wby]{callouts}
\usetikzlibrary{shapes.multipart}

\usepackage{array}
\usepackage{makecell}

\allowdisplaybreaks

\usepackage{amsthm}
\theoremstyle{definition}

\theoremstyle{remark}

\usepackage[utf8]{inputenc}
\usepackage[english]{babel}

\usepackage[caption=false,font=footnotesize]{subfig}
\usepackage[T1]{fontenc}
\usepackage{scalerel,stackengine}
\newcommand\reallywidecheck[1]{%
\savestack{\tmpbox}{\stretchto{%
  \scaleto{%
    \scalerel*[\widthof{\ensuremath{#1}}]{\kern-.6pt\bigwedge\kern-.6pt}%
    {\rule[-\textheight/2]{1ex}{\textheight}}%WIDTH-LIMITED BIG WEDGE
  }{\textheight}% 
}{0.5ex}}%
\stackon[1pt]{#1}{\scalebox{-1}{\tmpbox}}%
}
\stackMath
\IEEEoverridecommandlockouts
\IEEEoverridecommandlockouts

\definecolor{shadecolor}{RGB}{190,190,190}

\newcommand*{\mrn}{\textcolor{black}}

\newif\ifarxiv
\arxivtrue
\arxivfalse

\renewcommand{\baselinestretch}{.9}

\begin{document}

\title{\LARGE\bf
Leveraging Convex Relaxation to Identify the Feasibility of Conducting AC False Data Injection Attack in Power Systems}

%\author{Mohammadreza Iranpour$^{\ast}$, Mohammad Rasoul Narimani$^{\ast}$
%\thanks{${\ast}$: Department of Electrical and Computer Engineering, California State University Northridge (CSUN). Rasoul.narimani@csun.edu. Support from NSF contract \#2308498.}% 
%}

\author{\IEEEauthorblockN{1\textsuperscript{st} Mohammadreza Iranpour}
\IEEEauthorblockA{\textit{Department of Electrical and Computer Engineering} \\
\textit{California State University Northridge (CSUN)}\\
Los Angels, USA \\
\and
\IEEEauthorblockN{2\textsuperscript{nd} Mohammad Rasoul Narimani}
\IEEEauthorblockA{\textit{Department of Electrical and Computer Engineering} \\
\textit{California State University Northridge (CSUN)}\\
Los Angels, USA \\
Rasoul.narimani@csun.edu}}
\and

\thanks{ This work has been supported from NSF contract \#2308498.}%
}

\maketitle

\begin{abstract}
\mrn{FDI (False Data Injection) attacks are critical to address as they can compromise the integrity and reliability of data in cyber-physical systems, leading to potentially severe consequences in sectors such as power systems. The feasibility of FDI attacks has been extensively studied from various perspectives, including access to measurements and sensors, knowledge of the system, and design considerations using residual-based detection methods. Most research has focused on DC-based FDI attacks; however, designing AC FDI attacks involves solving a nonlinear optimization problem, presenting additional challenges in assessing their feasibility. Specifically, it is often unclear whether the infeasibility of some designed AC FDI attacks is due to the nonconvexity and nonlinearity inherent to AC power flows or if it stems from inherent infeasibility in specific cases, with local solvers returning infeasibility. This paper addresses this issue by leveraging the principle that if a convexified AC FDI attack design problem is infeasible, the attack design itself is infeasible, irrespective of nonlinear solution challenges. We propose an AC FDI attack design based on convexified power flow equations and assess the feasibility of the proposed attack by examining the extent of the attackable region. This approach utilizes a Quadratic Convex (QC) relaxation technique to convexify AC power flows. To evaluate the proposed method, we implement it on the IEEE 118-bus test system and assess the feasibility of an AC FDI attack across various attack zones.}
\end{abstract}

\section{Introduction}
\label{sec:Introduction}

\mrn{False Data Injection (FDI) attacks pose a significant threat to the integrity and reliability of modern power systems. These attacks involve injecting malicious data into the system's state estimation process, leading to incorrect operational decisions that can compromise the grid's stability and security. Numerous studies have reviewed FDI attacks, focusing on the requirements for conducting FDI attacks, construction methods, and detection and defense strategies. The construction methods have been examined from various perspectives, including FDI attacks with limited budgets, those based on state estimation with incomplete system knowledge, and data-driven approaches. In all of these cases, the feasibility of implementing a successful attack has been demonstrated and analyzed~\cite{zhang2019false,zhao2018generalized,zhang2018can}. In these cases, the attacker needs to create a plan that fits the specific circumstances and goals, making sure the attack remains undetected while achieving its objectives~\cite{boyaci2021joint,boyaci2022generating, boyaci2022infinite, boyaci2022spatio}.}

\mrn{While there has been a lot of research on DC-based FDI attacks, there is much less work on AC FDI attacks, and only a few studies have been published on this topic~\cite{jin2018power,rahman2014formal,teixeira2015secure,iranpour2024ac, iranpour2024designing}. The primary challenge in designing AC FDI attacks lies in the non-linear nature of AC power flow equations. Unlike DC power flow models, which are linear and easier to manipulate, AC models require solving non-linear optimization problems~\cite{iranpour2024ac}.}

\mrn{Although some research, such as~\cite{rahman2014formal}, has studied AC-based FDI attacks based on linearization around the target state under the assumption that SE is obtained, there are limited resources that have used convexification techniques to assess the feasibility of an AC FDI attack. For instance, in~\cite{jin2017semidefinite}, a novel convexification framework based on semidefinite programming (SDP) is employed to delineate the ``attackable region'' for any given set of measurement types and grid topology.}  

\mrn{In general, these problems involve non-linear variables, making it difficult to determine the feasibility of an attack. Specifically, it is often unclear whether an infeasible attack is due to the inherent challenges of solving a non-linear optimization problem or the infeasibility of the attack design itself. This hidden factor may hinder the successful design of an attack, a consideration that has not been addressed in recent studies. To address these challenges, this paper leverages a critical insight: if a convexified AC FDI attack is infeasible, the attack design itself is inherently infeasible, regardless of the non-linear nature of the problem~\cite{narimani2023tightening,narimani2020tightening, narimani2020strengthening,narimani2018comparison, narimani2018empirical,narimani2018improving}. This insight allows us to bypass the difficulties associated with non-linear optimization by focusing on the convexified version of the problem.}

\mrn{We propose a novel approach to designing AC FDI attacks based on convexified power flow equations. In fact, building on the work done in \cite{iranpour2024ac}, we have replaced the nonlinear power flow equations in the AC FDI attack process with convexified power flow equations. Many recent research efforts have developed convex relaxations of nonlinear power flow problems, as reviewed in \cite{molzahn2019survey}. Semidefinite programming, second-order cone programming, linear relaxation of power flow equations, and quadratic convex (QC) relaxation methods are examples of well-known approaches for convexifying nonlinear power flow equations. We use a Quadratic Convex (QC) relaxation technique to transform the nonlinear AC power flow equations into a convex form \cite{narimani2018improving}. This transformation simplifies the problem, making it easier to assess the feasibility of the attack design. If an attack designed in the convexified format is infeasible, it will certainly be infeasible in the nonlinear situation due to the design constraints. The Quadratic Convex (QC) relaxation is a promising approach that constructs convex envelopes around the trigonometric and product terms in the polar representation of the power flow equations \cite{coffrin2015qc}. In this paper, by defining these envelopes and also defining new corresponding linear variables for nonlinear variables, we have convexified the nonlinear power flow equations and then designed AC FDI attacks based on these equations.}

\mrn{This paper is organized as follows. In Section \ref{sec:optimal AC design}, we model an AC False Data Injection (FDI) attack based on the nonlinear power flow equations. Section \ref{sec:QC method} presents the convexification process of these nonlinear power flow equations using the Quadratic Convex (QC) relaxation method. In Section \ref{sec:efficiency assessment}, we analyze the efficiency of the proposed method in assessing the feasibility of AC FDI attacks by implementing it on the IEEE 118-bus test system. Finally, Section \ref{sec:conclusion} concludes the paper.}

\section{Optimal Designing of AC FDI attack}
\label{sec:optimal AC design}

\mrn{Designing an optimal AC false data injection (FDI) attack requires solving a challenging nonlinear optimization problem. Researchers have studied various methods for creating these optimal attacks. In this paper, we aim to optimize the magnitude of the attack vector injected into the targeted measurements using the unified approaches discussed in~\cite{iranpour2024ac}. }

\mrn{The metering process in modern power systems can be prone to errors, which may impact State Estimation (SE) operations. To identify these errors, the residual (the difference between observed measurements ($z$) and estimated values ($h(x)$)) is used as a detection parameter. If this residual exceeds a predefined threshold, it indicates the presence of bad data. This process is known as Bad Data Detection (BDD). However, if attackers inject a false data set ($c$) into these measurements in a way that satisfies equation~\eqref{eq:AC_residual}, and then use an attack vector $a$ based on equation~\eqref{eq:AC_undetectable}, they can manipulate the data without being detected by the system's detection modules.}

\vspace{-.5cm}
{\begin{align}
\label{eq:AC_residual}
&\nonumber r_a = z_a-h(x_a) \\
&\nonumber ~~~= z_a- h(x_a)+h(x)-h(x)\\
&~~~\nonumber= z+a-h(x_a)+h(x)-h(x)\\ &~~~= r+a- h(x_a)+h(x).
\end{align}}

\vspace{-.5cm}
{\begin{align}
\label{eq:AC_undetectable}
    a=h(x_a)-h(x).
\end{align}}
\vspace{-.5cm}

\mrn{In these equations, $r_a$ and $r$ are the residuals for under-attack and normal measurements, respectively. $x_a = x + c$ and $x$ are state vectors for under-attack and normal measurements, respectively. $h$ is the set of nonlinear power flow equations that relate measurements and states. Based on this assumption, the key to designing a successful attack vector $a$ in Equation \eqref{eq:AC_undetectable} is determining the function $h(x)$. By considering the following steps and assumptions, we can define this nonlinear function and then the attack vector $a$.}

\begin{enumerate}
    \item \mrn{\textbf{Rationalizing Changes in Power Flow Equations:}
    The attack must justify any changes in power flow equations resulting from the injected false data.}
    
    \item \mrn{\textbf{Defining Attacked and Normal Areas:}
  The attack area, where changes occur, should be delineated from the unaffected normal area. The attack area includes a contiguous set of buses, while the normal area remains unchanged. } 

    \item \mrn{\textbf{Maintaining Power Transfers:} During an FDI attack, the total power transferred between the affected and unaffected areas must stay the same. To achieve this, the attack area should be enclosed by buses that inject power. These buses manage power changes within the attack zone by creating specific injection measurements. Additionally, the state variables at the boundary of the attack area must stay the same to keep all changes confined within the attack zone. To define the attack zone, key focal buses have been chosen and the area has been expanded by including buses that inject no power and neighboring buses that inject power as the boundary.}

%this requires careful management of power injections at boundary buses to ensure all changes stay within the attack zone.

\item \mrn{\textbf{Algebraic Sum of Power in the Attack Zone:}
In the attack zone, the total power generated and consumed must remain unchanged. For buses that neither inject nor draw power, the sum of active and reactive power flows must be zero. For buses that do inject or draw power, the power injection after the attack equals the original injection power plus the changes in power flow in the lines connected to them within the attack zone.}
\end{enumerate}

\mrn{Continuing, we have modeled these assumptions as an optimization problem to calculate the attack vector $a$. Suppose the sets of buses and lines in the system are represented by $\mathcal{B}$, and $\mathcal{L}$, respectively. Also, $\mathcal{B_A}$, and $\mathcal{L_A}$ are the corresponding sets of buses and lines within the attack zone. Let $S_m = {P}_m + j {Q}_m$ represent the complex power injection, $V_m$ and $\theta_m$ represent the voltage magnitude and angle at bus~$m\in\mathcal{B_A}$, each line $\left(m,l\right)\in\mathcal{L_A}$ is modeled as a $\Pi$ circuit with mutual admittance $g_{ml}+j b_{ml}$ and shunt admittance $j b_{c,ml}$ and the voltage angle difference between buses $m$ and $l$ for $(m,l)\in\mathcal{L}$ is denoted as $\theta_{ml}=\theta_{m}-\theta_{l}$.
The difference between state variables, including both voltage magnitudes and angles, and their values before the FDI attack is shown in Equation~\eqref{eq:optimization} as vector $c$.}

\mrn{\begin{equation}
\label{eq:optimization}
 c= [\tilde{V}_{m}- V_{m,fix}, \tilde{\theta}_{m}-\theta_{m,fix}]
\end{equation}}

\mrn{Where $\tilde{V}_{m}$ and $\tilde{\theta}_{m}$ are the variables representing the voltage magnitude and angle that need to be calculated to conduct the attack, and $V_{m,fix}$ and $\theta_{m,fix}$ are the known values of voltage magnitude and angle before the attack. By minimizing the sum of squared differences between the variable states before and after the attack as the objective function, we can optimize the attack vector for designing the AC FDI attack problem, as represented in Equations~\eqref{eq:obj}-\eqref{eq:reactive_overload}.}

\begin{subequations}
\label{eq:optimization_problem}
\begin{align}
&\min\quad \textstyle\sum_{\small{{m}\in \mathcal{B_A}}}
 \left(\tilde{V}_{m}- V_{m,fix})^2 + (\tilde{\theta}_{m}-\theta_{m,fix})^2 \label{eq:obj} \right)\\
%&\quad\qquad\qquad+c_{1,k}\left( {\tilde P_k^g\cos (\psi_l ) - \tilde Q_k^g\sin (\psi_l )} \right)+c_{0,k}\\
%&\nonumber \min\quad \textstyle\sum_{\footnotesize{{k}\in \mathcal{G}}}
%c_{2,k}\left( {\tilde P_k^g\cos (\psi_l ) - \tilde Q_k^g\sin (\psi_l)} \label{eq:RQC obj} \right)^2\\
%&\quad\qquad\qquad+c_{1,k}\left( {\tilde P_k^g\cos (\psi_l ) - \tilde Q_k^g\sin (\psi_l )} \right)+c_{0,k}\\
&\nonumber \text{subject to} \quad \left(\forall i\in\mathcal{B_A}, \forall   \left(l,m\right) \in\mathcal{L_A}\right)\\
&\!\!\!\!\!\!\! g_{sh,i}\, \tilde{V}_i^2+\sum_{\substack{(l,m)\in \mathcal{L},\\\text{s.t.} \hspace{3pt} l=i}} \!\tilde{P}_{lm}+\!\!\sum_{\substack{(l,m)\in \mathcal{L},\\\text{s.t.} 
\label{eq:active_injection}\hspace{3pt} m=i}} \!\!\tilde{P}_{ml}= P_{i,G}-P_{i,D}, \\
&\!\!\!\!\!\!\! -b_{sh,i}\, \tilde{V}_i^2+\!\!\!\!\!\!\sum_{\substack{(l,m)\in \mathcal{L},\\ \text{s.t.} \hspace{3pt} l=i}} \!\!\tilde{Q}_{lm}+\!\!\!\sum_{\substack{(l,m)\in \mathcal{L},\\ \text{s.t.}\label{eq:reactive_injection}\hspace{3pt} m=i}} \!\!\!\!\tilde{Q}_{ml}=Q_{i,G}-Q_{i,D},\\
&\nonumber\!\!\!\!\!\!\! \tilde{P}_{lm} \!=\! g_{lm} \tilde{V}_l^2\! -\! g_{lm} \tilde{V}_l V_m\cos\left(\tilde{\theta}_{l}-\theta_{m}\right)\!\\
&\qquad\qquad -\! b_{lm} \tilde{V}_l V_m\sin\left(\tilde{\theta}_{l}-\theta_{m}\right),\\
\label{eq:qik1}
&\!\!\!\!\!\!\! \nonumber \tilde{Q}_{lm} = -\left(b_{lm}+b_{c,lm}/2\right) \tilde{V}_l^2 + b_{lm} \tilde{V}_l V_m\cos\left(\tilde{\theta}_{l}-\theta_{m}\right)\\ &\qquad\qquad  - g_{lm} \tilde{V}_l V_m\sin\left(\tilde{\theta}_{l}-\theta_{m}\right),\\
\label{eq:pki1}
&\nonumber \!\!\!\!\!\!\!\tilde{P}_{ml}\! =\! g_{lm} V_m^2\! -\! g_{lm} \tilde{V}_l V_m\cos\left(\tilde{\theta}_{l}-\theta_{m}\right)\!\\
&\qquad\qquad +\! b_{lm} \tilde{V}_l V_m\sin\left(\tilde{\theta}_{l}-\theta_{m}\right),\\
\label{eq:qki1}
&\!\!\!\!\!\!\!\nonumber \tilde{Q}_{ml} = -\left(b_{lm}+b_{c,lm}/2\right) V_m^2 + b_{lm} \tilde{V}_l V_m\cos\left(\tilde{\theta}_{l}-\theta_{m}\right)\\ &\qquad\qquad  + g_{lm} \tilde{V}_l V_m\sin\left(\tilde{\theta}_{l}-\theta_{m}\right)\\
\label{eq:active_overload}
&\!\!\!\!\!\!\! \tilde{P}_{ml} = W_{lm}*{P}_{lm}^{P
F},\\
\label{eq:reactive_overload}
&\!\!\!\!\!\!\! \tilde{Q}_{ml} = W_{lm}*{Q}_{lm}^{P
F}.
\end{align}
\end{subequations}

\mrn{In these equations $\tilde{P}_{ml}$, $\tilde{Q}_{ml}$, $\tilde{V}_{l}(\tilde{V}_{m})$ and $\tilde{\theta}_{l}(\tilde{\theta}_{m})$ represent the active and reactive power flow between buses $m$ and $l$ $(m,l)\in\mathcal{L}$, and the voltage magnitude and angle values of buses $m$ and $l$, after the attack. Similarly, $P_{ml}$, $Q_{ml}$, $V_{l}(V_{m})$ and $\theta_{l}(\theta_{m})$ represent the same quantities before the attack. Additionally, $P_{m,G}$, $Q_{m,G}$, $P_{m,D}$, and $Q_{m,D}$ are the active and reactive power generation and demand at bus $m$, respectively. It is notable that in Equations~\eqref{eq:active_injection} and~\eqref{eq:reactive_injection}, when these equations are written for zero-injection buses, the right-hand side of these equations is equal to zero. Equations \ref{eq:active_overload} and \ref{eq:reactive_overload} are additional constraints for overloading a specified line in the attack zone by a predefined coefficient $W$. By considering this constraint, we can design an optimal AC FDI attack for specific goals, such as overloading a certain line by a predefined coefficient. ${P}_{lm}^{P
F}$ and ${Q}_{lm}^{P
F}$ are the active and reactive power flows before the attack. After solving this optimization problem, we can calculate the power injection values for non-zero injection buses within the attack zone as follows:}

\mrn{\begin{subequations}
\begin{small}
\begin{align}
\label{eq:non_zero_injection}
&\tilde{P}_{m}=P_{m}+\sum_{(m,l)\in \mathcal{L}_A}(\tilde{P}_{m,l}-P_{m,l}),\\
&\tilde{Q}_{m}=Q_{m}+\sum_{(m.l)\in \mathcal{L_A}}(\tilde{Q}_{m,l}-Q_{m,l}).
\end{align}
\end{small}
\end{subequations}}

\mrn{Here, $\tilde{P_{m}}$, $\tilde{Q_{m}}$, $P_{m}$, and $Q_{m}$ represent the active and reactive power injections at bus $m$ after and before the attack, respectively. Notably, in Equations~\eqref{eq:non_zero_injection}, $\tilde{P}{m,l}$ and $\tilde{Q}{m,l} \in \mathcal{L}_A$ should be considered only for the lines within the attack zone that are connected to bus $m$. After calculating all of these values, we can define the attack vector $a$ based on Equation~\eqref{eq:AC_undetectable}. The attack vector $a$ is determined by the difference between the calculated values of power injections and their corresponding values before the attack, ensuring that the designed attack effectively meets the specified objectives. By accurately defining and optimizing $a$, we can achieve a successful AC FDI attack while adhering to the constraints and goals set for the attack.}

\mrn{\begin{small}
\begin{align}
\label{eq:attack_vector_final}
 &\nonumber a= [\tilde{P_{ml}}- P_{ml}, 
     \tilde{Q_{ml}}- Q_{ml},
     \tilde{P_{m}}- P_{m},\\
  &~~~~~~\nonumber   \tilde{Q_{m}}- Q_{m},
     \tilde{V_{m}}\angle \tilde{\theta_{m}}- V_{m}\angle\theta_{m}]^T
\end{align}
\end{small}}
\vspace{-.5cm}

\section{The QC Relaxation of the Proposed Problem}
\label{sec:QC method}
\mrn{As mentioned in Section \ref{sec:Introduction},  the quadratic convex (QC) relaxation \cite{coffrin2015qc} is a promising approach that utilizes convex envelopes around non-convex terms, including trigonometric functions, squared terms, and bilinear products. This method transforms the original non-convex problem into a convex one, which is generally easier to solve. The effectiveness of QC relaxation largely depends on the size and tightness of the variable bounds. By minimizing a linear objective function over the convex region defined by these convex envelopes, QC relaxation provides solutions that are either exact or close approximations to the original problem.
This technique has demonstrated significant success in solving or approximating solutions for many practical problems, particularly those that are NP-hard and challenging to tackle with traditional methods~\cite{NarimaniTPS, Narimani_GlobalSIP, narimani2020strengthening}. The approach has been applied to various fields, including operations research, engineering, and economics, showcasing its versatility and robustness. For further details and comprehensive overviews of QC relaxation and its diverse applications, please refer to the cited reference~\cite{molzahn2019survey}.}

\mrn{The QC relaxation is formed by introducing new variables 
$w_{ii}$, $w_{lm}$, $c_{lm}$, and $s_{lm}$ to represent the products of voltage magnitudes and trilinear monomials. These new variables capture the interactions between voltage magnitudes and trigonometric functions for connected buses. Specifically, $w_{ii}$ represents the squared terms of voltage magnitudes, $w_{lm}$ captures the product of voltage magnitudes at different buses, $c_{lm}$ denotes the bilinear terms involving voltage magnitudes and trigonometric functions, and 
$s_{lm}$ corresponds to additional trilinear products involving these functions. Equation~\eqref{eq:cs} mathematically represents these terms. By defining these variables, the QC relaxation transforms the original non-convex problem into a convex one, making it more tractable to solve while providing useful approximations or exact solutions for complex optimization challenges. }

\vspace{-.5cm}
\begin{subequations}
\label{eq:cs}
\begin{align}
w_{ii} &= V_i^2,  & \forall i \in\mathcal{N}, \\
w_{lm} &= V_l V_m,  & \forall \left(l,m\right) \in\mathcal{L}, \\
c_{lm} & =  w_{lm} \cos\left(\theta_{lm} \right), & \forall \left(l,m\right) \in\mathcal{L}, \\
s_{lm} & = w_{lm}\sin\left(\theta_{lm} \right),  & \forall \left(l,m\right) \in\mathcal{L}.
\end{align}
\end{subequations}

\mrn{For each line $\left(l,m\right)\in\mathcal{L}$, these definitions imply the following relationships between the variables $w_{ll}$, $c_{lm}$, and $s_{lm}$:}
\begin{subequations}
\label{eq:cs_relationships}
\begin{align}
\label{eq:Jabr}
&c_{lm}^2+s_{lm}^2=w_{ll}w_{mm},\\
\label{eq:cs_relationships_c}
&c_{lm}=c_{ml}, \\
\label{eq:cs_relationships_s}
&s_{lm}=-s_{ml}
\end{align}
\end{subequations}

\mrn{The QC relaxation is formulated by enclosing the squared and bilinear product terms within convex envelopes, which are represented here as set-valued functions:}
\begin{subequations}
\label{eq:product_envelopes}
\begin{align}
\label{eq:squareenvelopes}
\langle x^2\rangle^T =
\begin{cases}
\widecheck{x}: \begin{cases}\check{x} \geq x^2,\\
\widecheck{x} \leq \left({\overline{x}+\underline{x}}\right) x-{\overline{x} \underline{x}}.\\
\end{cases}
\end{cases}\\
\label{eq:mccormick}
\langle {xy}\rangle^M  =
\begin{cases}
\widecheck{xy}:\begin{cases}
\widecheck{xy} \geq {\underline{x}} y+ {\underline{y}} x-{\underline{x} \underline{y}},\\
\widecheck{xy} \geq {\overline{x}} y+ {\overline{y}} x-{\overline{x} \overline{y}},\\
\widecheck{xy} \leq {\underline{x}} y+ {\overline{y}} x-{\underline{x}} {\overline{y}},\\
\widecheck{xy} \leq {\overline{x}} y+ {\underline{y}} x-{\overline{x} \underline{y}}.\\
\end{cases}
\end{cases}
\end{align}
\end{subequations}

\mrn{where $\widecheck{x}$ and $\widecheck{xy}$ are ``auxiliary'' variables representing the corresponding sets.
The envelope $\langle x^2\rangle^T$ denotes the convex hull of the squared function, while the so-called ``McCormick envelope'' $\langle {xy}\rangle^M$ represents the convex hull of a bilinear product, as discussed in~\cite{kocuk2016strong,coffrin2016strengthening,mccormick1976computability}.}
\mrn{Additionally, the QC relaxation includes the formulation of convex envelopes for trigonometric functions, specifically $\left\langle\sin\left(x\right) \right\rangle^S$ and $\left\langle\cos\left(x\right) \right\rangle^C$. These envelopes are designed to approximate the sine and cosine functions, respectively, by providing convex relaxations that simplify the optimization process. The envelopes $\left\langle\sin\left(x\right) \right\rangle^S$ and $\left\langle\cos\left(x\right) \right\rangle^C$ are constructed as follows:}
\begin{subequations}
\label{eq:convex_envelopes_sin&cos}
\begin{align}
%\begin{align*}
%\begin{split}
\label{eq:sine envelope}
\nonumber &\left\langle \sin(x)\right\rangle^S =\\
%\end{split}
&\quad\;\begin{cases}
\widecheck{S}:\begin{cases}
\widecheck{S}\leq\cos\left(\frac{x^m}{2}\right)\left(x-\frac{x^m}{2}\right)+\sin \left(\frac{x^m}{2}\right),\\
\widecheck{S}\geq\cos\left(\frac{x^m}{2}\right)\left(x+\frac{x^m}{2}\right)-\sin\left(\frac{x^m}{2}\right),\\
\widecheck{S}\geq\frac{\sin\left({\underline{x}}\right)-\sin\left(\overline{x}\right)}{{\underline{x}-\overline{x}}}\left(x-{\underline{x}}\right)+\sin\left({\underline{x}}\right) \text{if~} \underline{x}\geq0,\\
\widecheck{S}\leq\frac{\sin\left({\underline{x}}\right)-\sin\left({\overline{x}}\right)}{{\underline{x}-\overline{x}}}\left(x-{\underline{x}}\right)+\sin\left({\underline{x}}\right) \text{if~} {\overline{x}}\leq0.
\end{cases}
\end{cases}\\
\label{eq:cosine envelope}
&\nonumber\left\langle\cos(x)\right\rangle^C=\\
&\quad\;\begin{cases}
\widecheck{C}:\begin{cases}
\widecheck{C}\leq 1-\frac{1 -\cos\left({x^m}\right)}{\left(x^m\right)^2}x^2,\\
\widecheck{C}\geq\frac{\cos\left(\underline{x}\right)-\cos\left({\overline{x}}\right)}{{\underline{x}-\overline{x}}}\left(x-{\underline{x}}\right)+\cos\left({\underline{x}}\right).  
\end{cases}
\end{cases}
\end{align}
\end{subequations}
\mrn{where $x^m= \max(\left|\underline{x}\right|,\left|\overline{x}\right|)$. The auxiliary variables $\check{S}$ and $\check{C}$ again represent the corresponding set. For $-90^\circ < \underline{x} < \overline{x} < 90^\circ$, bounds on the sine and cosine functions are:}
%For $-90^\circ < \underline{x} < \overline{x} < 90^\circ$, bounds on the sine functions, $\underline{s}$, $\overline{s}$, and cosine functions, $\underline{c}$, $\overline{c}$, are
%
\begin{subequations}
\begin{align}
& \underline{s} = \sin\left(\underline{x}\right) \leq \sin(x) \leq \overline{s} = \sin\left(\overline{x}\right), \\
\nonumber & \underline{c} = \min\left(\cos(\underline{x}),\cos(\overline{x})\right) \leq \cos(x) \\ & \quad \leq \overline{c} \!=\! \begin{cases} \max\left(\cos(\underline{x}),\cos(\overline{x})\right),\; \text{if~} \mathrm{sign}\left(\underline{x}\right) \!=\!  \mathrm{sign}\left(\overline{x}\right), \\ 1, \text{~otherwise}. \end{cases}\raisetag{1em}
\end{align}
\end{subequations} 

\mrn{By substituting the squared, product, and trigonometric terms in Equations~\ref{eq:obj}--~\eqref{eq:qki1} with the new variables $w_{ii}$, $w_{lm}$, $c_{lm}$, and $s_{lm}$, we can reformulate these equations to incorporate the convex envelopes. This substitution allows us to express the original equations in terms of these new variables, leading to a more tractable optimization problem. The convexified equations are as follows:} 

\begin{subequations}
\label{eq:qc}
\begin{align}
&\min\quad \textstyle\sum_{\small{{m}\in \mathcal{B_A}}}
 \left(\tilde{V}_{m}- V_{m,fix})^2 + (\tilde{\theta}_{m}-\theta_{m,fix})^2  \right)\\
&\nonumber \text{subject to} \quad \left(\forall i\in\mathcal{B_A}, \forall   \left(l,m\right) \in\mathcal{L_A}\right)\\
\label{eq:qc_p}
& P_i^g-P_i^d = g_{sh,i}\, w_{ii}+\sum_{\substack{(l,m)\in \mathcal{L}\\ \text{s.t.} \hspace{3pt} l=i}} P_{lm}+\sum_{\substack{(l,m)\in \mathcal{L}\\ \text{s.t.} \hspace{3pt} m=i}} P_{ml}, \\
\label{eq:qc_q}
& Q_i^g-Q_i^d = -b_{sh,i}\, w_{ii}+\sum_{\substack{(l,m)\in \mathcal{L}\\ \text{s.t.} \hspace{3pt} l=i}} Q_{lm}+\sum_{\substack{(l,m)\in \mathcal{L}\\ \text{s.t.} \hspace{3pt} m=i}} Q_{ml},\\
%\label{eq:qc_p}
%&\qquad P_i^g-P_i^d =  \sum_{k\in \mathcal{N}} \mathbf{G}_{ik} c_{ik}  + \mathbf{B}_{ik} s_{ik}, \\
%\label{eq:qc_q}
%& \qquad Q_i^g-Q_i^d = \sum_{k\in \mathcal{N}} \mathbf{G}_{ik}s_{ik} -
%\mathbf{B}_{ik} c_{ik},\\[-3pt]
%\label{eq:qc_ref}
%& \theta_{ref}=0,\\
%\label{eq:qc_Pg}
%& P_i^{g,l}\leq P_i^g\leq P_i^{g,u},\\
%\label{eq:qc_Qg}
%& Q_i^{g,l}\leq Q_i^g\leq Q_i^{g,u},\\
\label{eq:qc_V}
&  (\underline{V}_i)^2\leq w_{ii} \leq (\overline{V}_i)^2,\\
%\label{eq:qc_theta}
%&\theta_{ik}^{l}\leq \theta_{ik}\leq \theta_{ik}^{u},\\
\label{eq:qc_pik}
& P_{lm} = g_{lm} w_{ll} - g_{lm} c_{lm} - b_{lm} s_{lm}, \\
\label{eq:qc_qik}
& Q_{lm} = -\left(b_{lm}+b_{sh,lm}/2\right) w_{ii} + b_{lm} c_{lm}- g_{lm} s_{lm}, \\
%\label{eq:qc_sik}
%& \left(P_{ik}\right)^2+\left(Q_{ik}\right)^2 \leq \left(S_{ik}^u\right)^2, \\
\label{eq:qc_wii}
&  w_{ii} \in\left\langle V_i^2 \right\rangle^T, \\
\label{eq:qc_wik}
&  w_{lm} \in \left\langle V_l V_m \right\rangle^M, \\
\label{eq:qc_cik}
&  c_{lm} \in \left\langle w_{lm}\left\langle\cos\left(\theta_{lm} \right)\right\rangle^C\right\rangle^M, \\
\label{eq:qc_sik}
& s_{lm} \in \left\langle w_{lm} \left\langle\sin\left(\theta_{lm} \right)\right\rangle^S\right\rangle^M, \\
\label{eq:qc_jabr}
%&  c_{lm}^2+s_{lm}^2 \leq w_{ll}\,w_{mm} \\
%\label{eq:qc_others}
&  \text{Equations~}\eqref{eq:active_overload}\text{--}\eqref{eq:reactive_overload},\,\eqref{eq:cs_relationships_c},\,\eqref{eq:cs_relationships_s}.
\end{align}
\end{subequations}

%Note that the non-convex constraint~\eqref{eq:Jabr} is relaxed to~\eqref{eq:qc_jabr} using a less-stringent rotated second-order cone constraint~\cite{jabr2006radial}. Also

\mrn{Note that the trilinear terms in~\eqref{eq:active_injection}--~\eqref{eq:qki1} are addressed in~\eqref{eq:qc_wik}--\eqref{eq:qc_sik} by recursively applying \mbox{McCormick} envelopes~\eqref{eq:mccormick} (i.e., first applying~\eqref{eq:mccormick} to the product of voltage magnitudes to obtain $w_{lm}$ and then to the product of $w_{lm}$ and $\left\langle\cos\left(\theta_{lm} \right)\right\rangle^C$ or $\left\langle\sin\left(\theta_{lm} \right)\right\rangle^S$). }
%\textcolor{red}{The rotated quadratic cone in~\eqref{eq:qc_jabr} can be formulated as SOCP constraint.}
\mrn{The optimization problem described in Equation~\eqref{eq:qc} involves a quadratic objective function with linear constraints. The objective function is represented as $\frac{1}{2} x^T Q x + c^T x + d$, where 
$Q$ is a symmetric matrix. When $Q$ is positive semidefinite (PSD), the quadratic function becomes convex. As a result, the problem can be classified as a convex quadratic program (QP). Convex QPs can be solved globally, ensuring that the solutions found are optimal. For convex QPs, various algorithms like interior-point methods, active-set methods, and gradient-based methods can be used. These methods are designed to find the global minimum efficiently.}

\mrn{Based on these relaxed power flow equations, we can design an AC FDI attack using the process outlined in Section~\ref{sec:optimal AC design}. It is important to note that, in this context, we need to fix all corresponding variables in their relaxed format, including those in Equation \ref{eq:cs} ($w_{ii}$, $w_{lm}$, $c_{lm}$, and $s_{lm}$) for the variables in the out-of-the-attack zone. Additionally, to ensure that all changes are confined within the attack zone, $w_{ii}$ should be fixed for the boundary buses in this new set of convexified equations.}

\section{Efficiency Assessment of Proposed Method }
\label{sec:efficiency assessment}

\mrn{In this section, the proposed approach for assessing the feasibility of conducting optimal AC-FDI attacks is applied to the IEEE 118-bus test system from the PGLib-OPF v18.08 benchmark library~\cite{pglib} to evaluate its effectiveness. We have implemented two different scenarios for this purpose. In the first scenario, we demonstrate whether an optimal AC-FDI attack can be feasible or infeasible by considering two different attack zones. In the second scenario, we show how varying the values of $W$ in constraint~\eqref{eq:active_overload} affects the feasibility of the optimal AC-FDI attack, as addressed in Equation~\ref{eq:qc}, within a specific zone.}

\begin{figure*}
    \centering
    \hspace{-0.8cm}
\captionsetup{justification=centering}
\includegraphics[scale=0.31,trim= 21cm 0.4cm 0.7cm .4cm,clip]{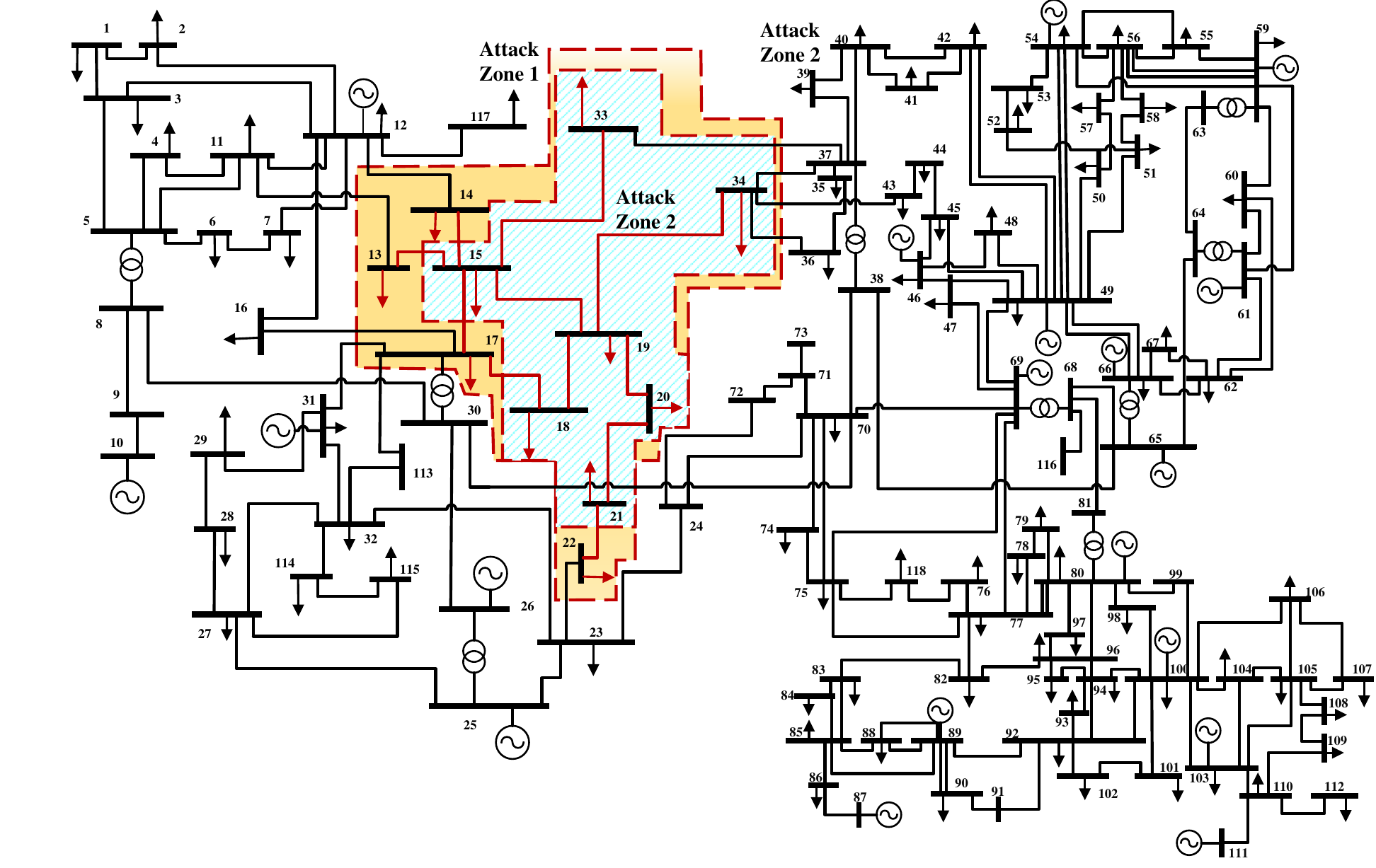}
	 \caption{{One-line diagram depicting the IEEE 118-bus test system, with the attack zones indicated.}}
	\label{fig:attack_zone1,2_118}
\end{figure*}

\vspace{.2cm}
\textbf{First Scenario}:  

\mrn{In this scenario, we considered two different attack zones and designed AC FDI attacks for those zones using the proposed methods in Sections~\ref{sec:optimal AC design} and~\ref{sec:QC method}. In this regard, as shown in Fig.~\ref{fig:attack_zone1,2_118}, we first consider buses $13,14,15,17,18,19,20,21,22,33$ and $34$ as attack zone 1. By implementing the design process based on the convexified power flow equations and setting $W=1.2$ in constraint \eqref{eq:active_overload} for the line between buses $18$ and $19$, the solver can find a feasible solution for the objective function. The corresponding values of voltage magnitudes and angles of buses in the attack zone before and after the attack are listed in Table~\ref{Table:Voltages_12}.}

\mrn{It is noteworthy that in Table \ref{Table:Voltages_12}, the voltages of boundary buses remain unchanged after the attack to prevent the spread of power transmission changes in the lines. However, the voltage magnitudes and angles of targeted buses in the attack zone have changed. In Table \ref{Table:Voltages_12}, gray cells correspond to the buses in the attack zone with variable values, while white cells indicate boundary buses with fixed voltage magnitudes and angles. Additionally, all power flow and power injection measurements in the lines and buses of the attack zone that need to be manipulated to implement a successful attack are shown in red in Fig.~\ref{fig:attack_zone1,2_118}.}

\begin{table}
\caption{Voltage magnitude and angle of busses in the first attack zone of first scenario, for before and after a feasible optimal attack.}
\label{Table:Voltages_12}
\begin{tabular}{|c|cc|cc|}
\hline
                                                                                 & \multicolumn{2}{c|}{\textbf{Before Attack}}                                                                                                                   & \multicolumn{2}{c|}{\textbf{Optimal Feasible Attack}}                                                                                                        \\ \cline{2-5} 
\multirow{-2}{*}{\textbf{\begin{tabular}[c]{@{}c@{}}Bus\\  Number\end{tabular}}} & \multicolumn{1}{c|}{\textbf{\begin{tabular}[c]{@{}c@{}}Voltage \\ Magnitude\end{tabular}}} & \textbf{\begin{tabular}[c]{@{}c@{}}Voltage\\ Angle\end{tabular}} & \multicolumn{1}{c|}{\textbf{\begin{tabular}[c]{@{}c@{}}Voltage\\ Magnitude\end{tabular}}} & \textbf{\begin{tabular}[c]{@{}c@{}}Voltage\\ Angle\end{tabular}} \\ \hline
13                                                                               & \multicolumn{1}{c|}{0.9683}                                                                & 11.6297                                                          & \multicolumn{1}{c|}{0.9683}                                                               & 11.6297                                                          \\ \hline
14                                                                               & \multicolumn{1}{c|}{0.9836}                                                                & 11.7715                                                          & \multicolumn{1}{c|}{0.9836}                                                               & 11.7715                                                          \\ \hline
\rowcolor[HTML]{C0C0C0} 
15                                                                               & \multicolumn{1}{c|}{\cellcolor[HTML]{C0C0C0}0.97}                                          & 11.4741                                                          & \multicolumn{1}{c|}{\cellcolor[HTML]{C0C0C0}0.9707}                                       & 11.4594                                                          \\ \hline
17                                                                               & \multicolumn{1}{c|}{0.9951}                                                                & 13.9952                                                          & \multicolumn{1}{c|}{0.9951}                                                               & 13.9952                                                          \\ \hline
\rowcolor[HTML]{C0C0C0} 
18                                                                               & \multicolumn{1}{c|}{\cellcolor[HTML]{C0C0C0}0.973}                                         & 11.7808                                                          & \multicolumn{1}{c|}{\cellcolor[HTML]{C0C0C0}0.9956}                                       & 11.763                                                           \\ \hline
\rowcolor[HTML]{C0C0C0} 
19                                                                               & \multicolumn{1}{c|}{\cellcolor[HTML]{C0C0C0}0.962}                                         & 11.3146                                                          & \multicolumn{1}{c|}{\cellcolor[HTML]{C0C0C0}0.946}                                        & 11.3034                                                          \\ \hline
\rowcolor[HTML]{C0C0C0} 
20                                                                               & \multicolumn{1}{c|}{\cellcolor[HTML]{C0C0C0}0.9569}                                        & 12.191                                                           & \multicolumn{1}{c|}{\cellcolor[HTML]{C0C0C0}0.9578}                                       & 12.1781                                                          \\ \hline
\rowcolor[HTML]{C0C0C0} 
21                                                                               & \multicolumn{1}{c|}{\cellcolor[HTML]{C0C0C0}0.9577}                                        & 13.778                                                           & \multicolumn{1}{c|}{\cellcolor[HTML]{C0C0C0}0.9586}                                       & 13.7611                                                          \\ \hline
22                                                                               & \multicolumn{1}{c|}{0.969}                                                                 & 16.3316                                                          & \multicolumn{1}{c|}{0.969}                                                                & 16.3316                                                          \\ \hline
33                                                                               & \multicolumn{1}{c|}{0.9709}                                                                & 10.8538                                                          & \multicolumn{1}{c|}{0.9709}                                                               & 10.8538                                                          \\ \hline
34                                                                               & \multicolumn{1}{c|}{0.984}                                                                 & 11.5114                                                          & \multicolumn{1}{c|}{0.984}                                                                & 11.5114                                                          \\ \hline
\end{tabular}
\end{table}

\mrn{Furthermore, when considering buses $15,18,19,20,21,33$ and $34$ as the second attack zone, as shown in Fig. \ref{fig:attack_zone1,2_118}, the solver fails to find a feasible solution for the objective function. This indicates that the corresponding QP problem, the optimization problem in~\eqref{eq:qc}, and consequently there is a guarantee that the nonconvex nonlinear optimization problem in~\eqref{eq:optimization_problem} has no feasible solution, which is a crucial point in designing AC FDI attack. While the aim of designing a sparse attack vector is to target zones with fewer measurements, it's important to note that making the region smaller can render the solution infeasible. Therefore, a trade-off must be struck between minimizing the area and ensuring solution feasibility.}

%Furthermore, when considering buses $15,18,19,20,21,33$ and $34$ as the second attack zone, as it has been shown in the Fig.\ref{fig:attack_zone1,2_118} the solver fails to find a feasible solution for the objective function. This indicates that the corresponding non-linear problem has no feasible solution. This is a crucial point in designing a sparse AC FDI attack. While designing a sparse attack vector aims to target zones with fewer measurements, it's important to note that making the region smaller can render the solution infeasible. Therefore, a trade-off must be struck between minimizing the area and ensuring solution feasibility.

\begin{figure*}
    \centering
    \hspace{-0.3cm}
\captionsetup{justification=centering}
\includegraphics[scale=0.3,trim= 22cm 0.4cm 0.7cm .4cm,clip]{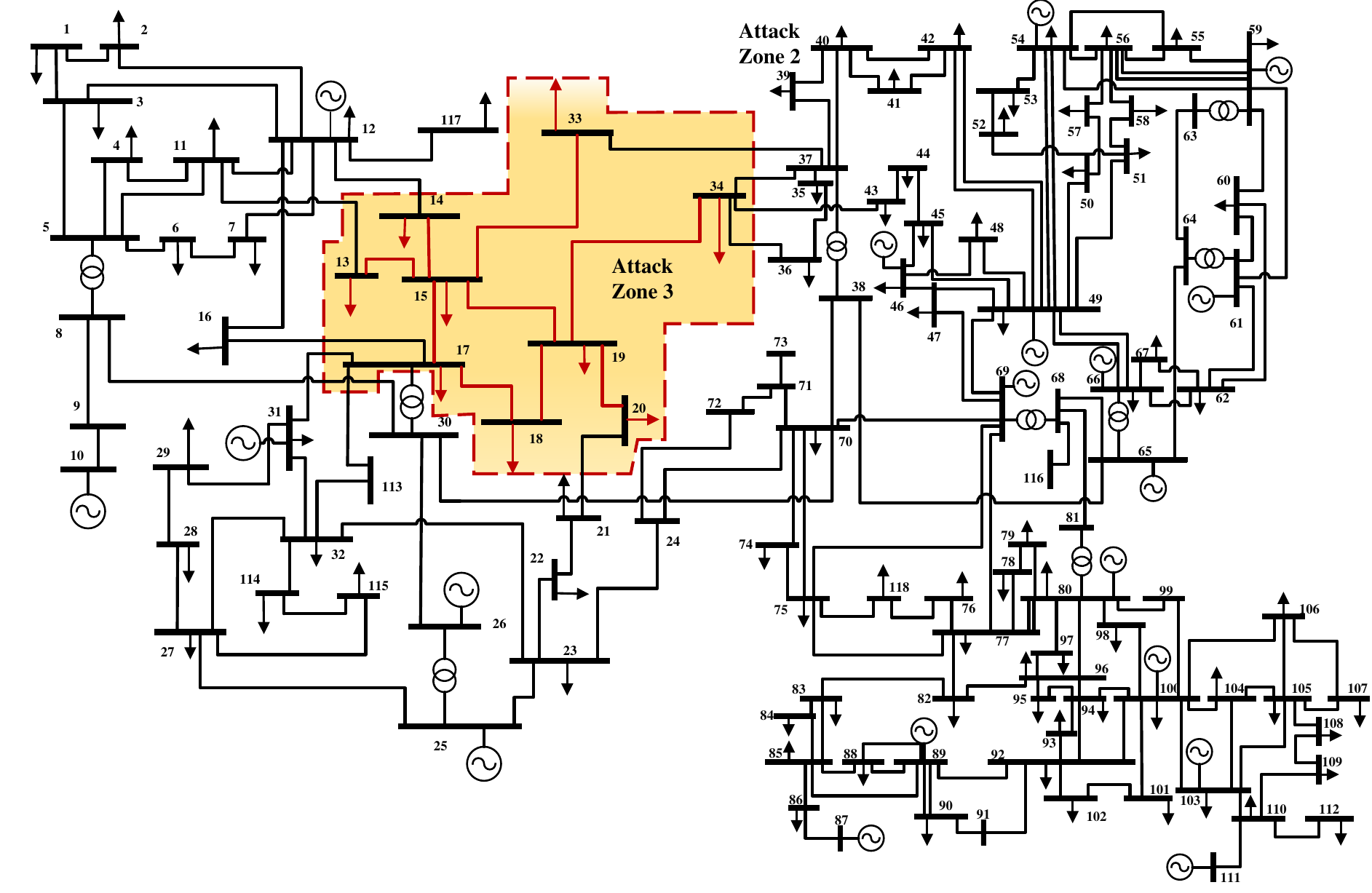}
	 \caption{Third attack zone in One-line diagram of the IEEE 118-bus test system.}
	\label{fig:attack_zone3_118}
\end{figure*}

\vspace{.2cm}
\textbf{ Second Scenario }:

\mrn{In this scenario, we considered buses $15,18,19,20,33,34$ and $36$ as the third attack zone as it is shown in~Fig.\ref{fig:attack_zone3_118}. If we enforce constraint~\eqref{eq:active_overload} for the line between buses $18$ and $19$ with $W=1.2$, the solver can find a feasible solution for Equation~\eqref{eq:qc}. The corresponding values of voltage magnitudes and angles of buses in the attack zone, before and after the attack implementation, are listed in Table \ref{Table:Voltages_3}. In this table, gray cells correspond to the buses in the attack zone with variable values, while white cells indicate boundary buses with fixed voltage magnitudes and angles. 
By enforcing power flow in the line that connects buses $18$ and $19$, i.e., setting $W=2$ in Equation~\ref{eq:active_overload}, the solver fails to find a feasible solution for the relaxed problem. Consequently, there will be no solution for the original optimization problem in~\eqref{eq:optimization_problem}.
Moreover, this demonstrates that modifying the constraints of the proposed optimization problem can lead to infeasibility. Thus, both the size of the attack area and the constraints can contribute to the infeasibility of the optimal AC FDI attack.
If the convex problem in~\eqref{eq:qc} is infeasible then the corresponding nonlinear problem solution is certainly infeasible because the solver cannot find a feasible solution for the proposed problem. By convexifying the design of the AC FDI attack problem, we can easily check if the problem is feasible and has a solution, unlike the nonconvex problem which cannot be solved using current local solvers even if a solution exists. Thus, the proposed method can be exploited to ensure the feasibility of conducting an AC FDI attack.}

\begin{table}
\caption{Voltage magnitude and angle of busses in the third attack zone of the second scenario, for before and after feasible optimal attack situations.}
\label{Table:Voltages_3}
\begin{tabular}{|c|cc|cc|}
\hline
                                                                                 & \multicolumn{2}{c|}{\textbf{Before attack}}                                                                                                                   & \multicolumn{2}{c|}{\textbf{Optimal feasible attack}}                                                                                                        \\ \cline{2-5} 
\multirow{-2}{*}{\textbf{\begin{tabular}[c]{@{}c@{}}Bus \\ Number\end{tabular}}} & \multicolumn{1}{c|}{\textbf{\begin{tabular}[c]{@{}c@{}}Voltage \\ Magnitude\end{tabular}}} & \textbf{\begin{tabular}[c]{@{}c@{}}Voltage\\ Angle\end{tabular}} & \multicolumn{1}{c|}{\textbf{\begin{tabular}[c]{@{}c@{}}Voltage\\ Magnitude\end{tabular}}} & \textbf{\begin{tabular}[c]{@{}c@{}}Voltage\\ Angle\end{tabular}} \\ \hline
13                                                                               & \multicolumn{1}{c|}{0.9683}                                                                & 11.6297                                                          & \multicolumn{1}{c|}{0.9683}                                                               & 11.6297                                                          \\ \hline
14                                                                               & \multicolumn{1}{c|}{0.9836}                                                                & 11.7715                                                          & \multicolumn{1}{c|}{0.9836}                                                               & 11.7715                                                          \\ \hline
\rowcolor[HTML]{C0C0C0} 
15                                                                               & \multicolumn{1}{c|}{\cellcolor[HTML]{C0C0C0}0.97}                                          & 11.4741                                                          & \multicolumn{1}{c|}{\cellcolor[HTML]{C0C0C0}0.9477}                                       & 11.4456                                                          \\ \hline
17                                                                               & \multicolumn{1}{c|}{0.9951}                                                                & 13.9952                                                          & \multicolumn{1}{c|}{0.9951}                                                               & 13.9952                                                          \\ \hline
\rowcolor[HTML]{C0C0C0} 
18                                                                               & \multicolumn{1}{c|}{\cellcolor[HTML]{C0C0C0}0.973}                                         & 11.7808                                                          & \multicolumn{1}{c|}{\cellcolor[HTML]{C0C0C0}0.9704}                                       & 11.7238                                                          \\ \hline
\rowcolor[HTML]{C0C0C0} 
19                                                                               & \multicolumn{1}{c|}{\cellcolor[HTML]{C0C0C0}0.962}                                         & 11.3146                                                          & \multicolumn{1}{c|}{\cellcolor[HTML]{C0C0C0}0.9615}                                       & 11.2656                                                          \\ \hline
20                                                                               & \multicolumn{1}{c|}{0.9569}                                                                & 12.191                                                           & \multicolumn{1}{c|}{0.9569}                                                               & 12.191                                                           \\ \hline
33                                                                               & \multicolumn{1}{c|}{0.9709}                                                                & 10.8538                                                          & \multicolumn{1}{c|}{0.9709}                                                               & 10.8538                                                          \\ \hline
34                                                                               & \multicolumn{1}{c|}{0.984}                                                                 & 11.5114                                                          & \multicolumn{1}{c|}{0.984}                                                                & 11.5114                                                          \\ \hline
\end{tabular}
\end{table}

\section{conclusion }
\label{sec:conclusion}

\mrn{This paper addresses the challenges in assessing the feasibility of designing AC False Data Injection (FDI) attacks. While most existing research has focused on DC-based FDI attacks, AC FDI attacks present additional complexities due to the need to solve nonlinear optimization problems. This study highlights the difficulty in distinguishing between the infeasibility of an attack due to the nonlinearity and nonconvexity associated with the AC FDI attack design problem and the inherent infeasibility of the attack design itself.}
\mrn{To overcome these challenges, we proposed a method that utilizes convexified power flow equations to design AC FDI attacks. By employing a Quadratic Convex (QC) relaxation technique, we were able to convexify the AC power flows and thereby provide a more straightforward assessment of the attack's feasibility. Our approach was validated on the IEEE 118-bus test system, demonstrating its effectiveness in determining the feasibility of optimal AC FDI attacks across different attack zones and various constraints for the objective function.}
\mrn{In addition to demonstrating the efficiency of the proposed QC relaxation method in assessing the feasibility of AC FDI attacks, the findings in this paper highlight the importance of considering various factors that influence the feasibility of such attacks. These factors include the attack area and the constraints of the optimization problem. This research offers a robust framework for future studies to design optimal AC FDI attacks, addressing perspectives such as sparse AC FDI attacks. Unlike sparse DC FDI attacks, which have been extensively studied, there is limited research on sparse AC FDI attacks. This paper aims to fill that gap and provide a foundation for further exploration in this area.}

\bibliographystyle{IEEEtran}
\IEEEtriggeratref{40}
\bibliography{ref}
\end{document}